\documentclass{webofc}
\usepackage[varg]{txfonts}   
\usepackage{siunitx}
\usepackage[]{cleveref}
\usepackage{xcolor} \usepackage{ulem} \usepackage{changebar}

\colorlet{ins}{blue} \colorlet{del}{red}

\usepackage[]{subcaption}
\usepackage{amsmath}
\usepackage{lineno}
\usepackage[
    protrusion=true
]{microtype}
\usepackage{everypage}
\usepackage{afterpage}

\newcommand{\pT}{\ensuremath{p_\mathrm{T}}}
\newcommand{\ktwos}{\ensuremath{\mathrm{\kappa}_2/\langle~\mathrm{p}~+~\overline{\mathrm{p}}~\rangle}}

\newcommand{\kthreektwo}{\ensuremath{\mathrm{\kappa}_3/\mathrm{\kappa}_2}}
\newcommand{\kfourktwo}{\ensuremath{\mathrm{\kappa}_4/\mathrm{\kappa}_2}}
\newcommand{\ksixktwo}{\ensuremath{\mathrm{\kappa}_6/\mathrm{\kappa}_2}}
\newcommand{\dNchdeta}{\ensuremath{\langle \mathrm{d}N_\mathrm{ch}/\mathrm{d}\eta \rangle_{\mathrm{|\eta| < 0.5}}}}

\newcommand{\pr}{\ensuremath{\mathrm{p}}}
\newcommand{\apr}{\ensuremath{\overline{\pr}}}

\newcommand{\sqrts}{\ensuremath{\sqrt{s}}}
\newcommand{\sqrtsNN}{\ensuremath{\sqrt{s_\mathrm{NN}}}}

\DeclareSIUnit \clight {\text{\ensuremath {c}}}
\newcommand{\GeVc}{\giga \electronvolt \per \clight}
\newcommand{\GeV}{\giga \electronvolt}
\newcommand{\MeV}{\mega \electronvolt}
\newcommand{\TeV}{\tera \electronvolt}

\begin{document}
\title{
Higher-Order Fluctuations: Unveiling the Final Frontier of QCD at the LHC with ALICE
}
%
%

\author{
    \firstname{Ilya} \lastname{Fokin}\inst{1}\fnsep\thanks{\email{fokin@physi.uni-heidelberg.de}}, on behalf of the ALICE Collaboration
}

\institute{
    Physikalisches Institut, Universität Heidelberg
}



\abstract{%
Lattice QCD (LQCD) calculations predict that chiral symmetry is restored in a smooth crossover transition between a quark--gluon plasma and a hadron resonance gas (HRG) at vanishing net-baryon density, a condition realized in heavy-ion collisions at the LHC.
In this regime, the net-baryon number cumulants computed using the HRG and LQCD partition functions are in good agreement up to third order.
However, starting with the fourth-order cumulants, the LQCD results are significantly lower than the corresponding HRG results.
This offers a unique opportunity to experimentally verify the full QCD partition function by measuring the fourth-order cumulants of the net-proton number distributions.
We present net-proton number cumulants up to sixth order in proton--proton collisions at top LHC energy recorded by ALICE to search for effects of a possible chiral phase transition in this small system.
An extension of the net-proton measurements in Pb--Pb collisions to fourth order is also presented.
In addition to providing experimental access to the full QCD partition function, these measurements will, for the first time, allow to distinguish between different mechanisms of baryon production.

}
\maketitle

\section{Introduction}
\label{intro}

Fluctuations of conserved charges are an important tool to study the nature of the chiral phase transition in QCD.
For two massless quarks and vanishing baryon chemical potential $\mu_\mathrm{B}$, the phase transition is expected to be of second order belonging to the O(4) universality class \cite{friman_fluctuations_2011-1}.
Finite physical quark masses lead to a smooth crossover with a pseudocritical temperature $T_\mathrm{pc} = \SI{156.5}{\MeV}$ in LQCD calculations \cite{hotqcd_collaboration_skewness_2020}.
Experimental support for the phase transition comes from the analysis of measured hadron multiplicities from Pb--Pb collisions at LHC energies \cite{andronic_decoding_2018}.
Susceptibilities $\chi_n^X$ for a conserved charge $X$ (e.g. the baryon number $\mathrm{B}$), which can be calculated in LQCD can be related to cumulants of the conserved charge $\kappa_n(X)$ via
\begin{equation}
    \chi_n^X = \frac{1}{VT^3} \frac{\partial^n \ln Z}{\partial (\mu_X/T)^n} = \frac{1}{VT^3} \kappa_n(X),
\end{equation}
where $V$ is the volume, $T$ is the temperature and $Z$ is the grand canonical partition function.
Cumulants provide an alternative description of probability distributions to moments.
Up to fourth order, they are given by $\kappa_1(X) = \langle X \rangle$, $\kappa_2(X) = \langle(X - \langle X \rangle)^2 \rangle$, $\kappa_3(X) = \langle(X - \langle X \rangle)^3 \rangle$ and $\kappa_4(X) = \langle(X - \langle X \rangle)^4 \rangle - 3 \left(\kappa_2(X)\right)^2$, where $\langle \cdot \rangle$ denotes the event-by-event average.
Using ratios of cumulants to cancel the factor $VT^3$ to first order, it is possible to experimentally test these LQCD results in heavy-ion collisions of sufficiently high energy, where $\mu_\mathrm{B}$ is small.
In particular, the full LQCD calculation predicts \cite{hotqcd_collaboration_skewness_2020} $\chi_4^\mathrm{B}/\chi_2^\mathrm{B} \approx 0.75$ and even a negative $\chi_6^\mathrm{B}/\chi_2^\mathrm{B}$ whereas a HRG model with Poissonian yield fluctuations would give unity for both ratios.
The STAR Collaboration has measured net-proton \ksixktwo{} consistent with zero in pp collisions at the highest multiplicities, hinting at a possible sign change \cite{abdulhamid_measurements_2024}.
Due to the small masses of up and down quarks and the proximity to the chiral limit \cite{hotqcd_collaboration_skewness_2020}, it remains an open question whether remnants of the second order phase transition can also be observed at LHC energies.

A direct comparison of cumulant ratios and LQCD is not trivial.
While LQCD describes an infinite system in thermal equilibrium, cumulants are measured in highly dynamic heavy-ion collisions in finite momentum and rapidity acceptances.
The net-proton number is usually used as a proxy for the baryon number \cite{kitazawa_relation_2012}.
Local baryon number conservation can modify the measured fluctuations due to additional correlations \cite{braun-munzinger_imprint_2024}, which have to be considered in the calculation of the baseline.
Volume fluctuations modify the cumulants because the number of particle sources is not necessarily identical for two collisions with similar multiplicities in the final state \cite{braun-munzinger_bridging_2017}.

\section{Analysis strategy}
\label{analysis}

This contribution presents ratios of net-proton cumulants obtained from data recorded by ALICE in the 2018 pp and Pb--Pb data taking periods at $\sqrts = \SI{13}{\TeV}$ and $\sqrtsNN = \SI{5.02}{\TeV}$ respectively. ALICE and its subdetectors are described in \cite{Collaboration2008ALICE}.
The V0A and V0C scintillator arrays are used for multiplicity (pp collisions) and centrality (Pb--Pb collisions) estimation.
Protons are identified using the specific energy loss in the Inner Tracking System (ITS) and the Time Projection Chamber (TPC).
In pp collisions, Time-Of-Flight (TOF) information is used in addition for transverse momenta $\pT \geq \SI{1}{\GeV}$.
The moments of the reconstructed net-proton number are then corrected for efficiency using a binomial model \cite{nonaka_more_2017, luo_efficiency_2019}.

\section{Net-proton fluctuations in proton-proton collisions}
\label{ppcollisions}

\begin{figure}[]
    \centering
    \begin{subfigure}{0.32\textwidth}
        \centering
        \includegraphics[width=\textwidth]{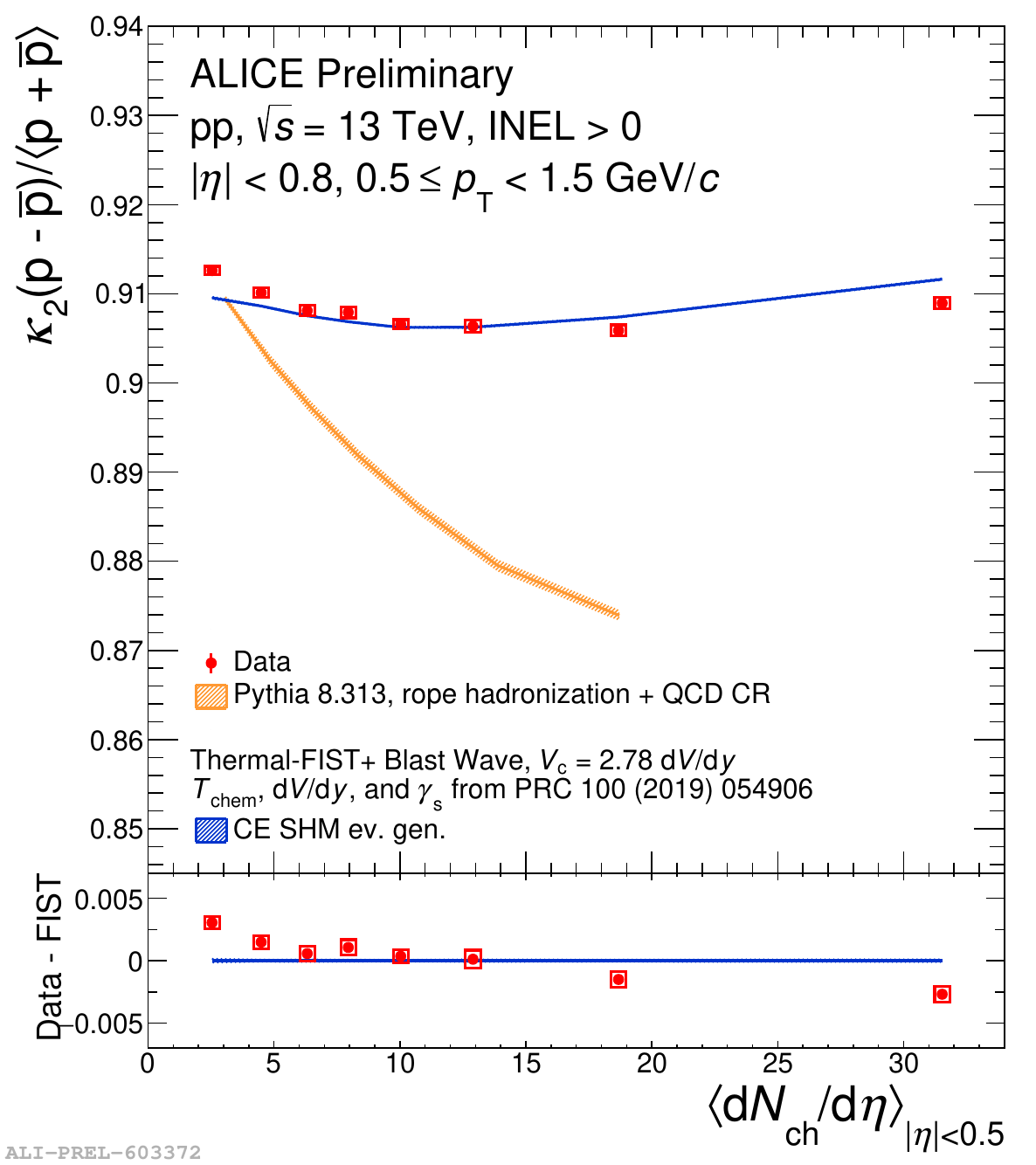}
        \label{fig:pp_k2}
    \end{subfigure}
    \hfill
    \begin{subfigure}{0.32\textwidth}
        \centering
        \includegraphics[width=\textwidth]{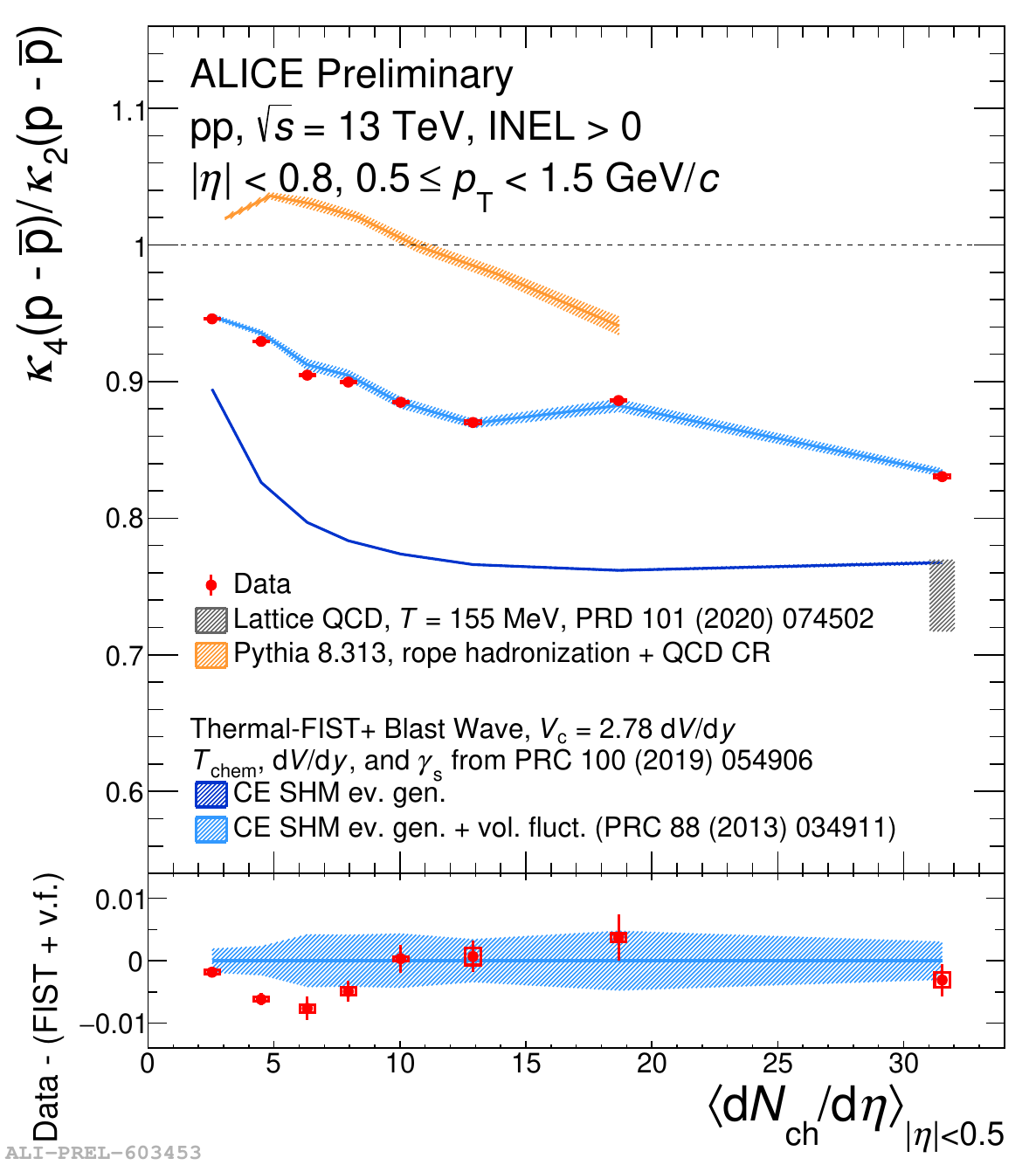}
        \label{fig:pp_k4k2}
    \end{subfigure}
    \hfill
    \begin{subfigure}{0.32\textwidth}
        \centering
        \includegraphics[width=\textwidth]{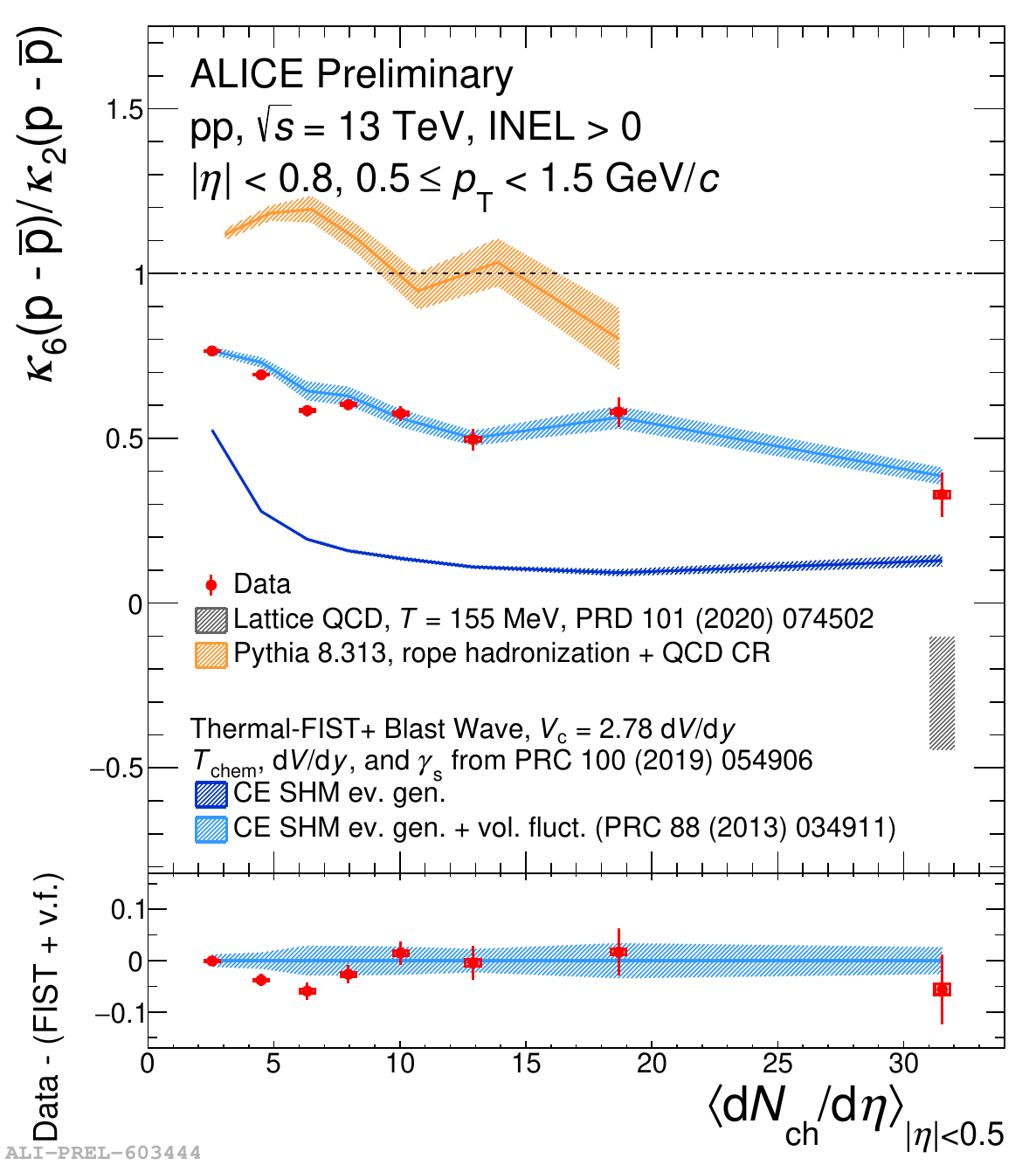}
        \label{fig:pp_k6k2}
    \end{subfigure}
    \caption{Net-proton number cumulant ratios $\kappa_2(\pr - \apr)/\langle \pr + \apr \rangle$ (left), $\kappa_4(\pr - \apr)/\kappa_2(\pr - \apr)$ (center) and $\kappa_6(\pr - \apr)/\kappa_2(\pr - \apr)$ (right) in pp collisions at $\SI{13}{\TeV}$ as a function of the charged particle multiplicity density at midrapidity $\dNchdeta$.}
    \label{fig:pp_cumulants}
\end{figure}

\Cref{fig:pp_cumulants} shows net-proton cumulant ratios up to sixth order in pp collisions as a function of the charged-particle multiplicity density at midrapidity \dNchdeta{} in the pseudorapidity $\left| \eta \right|~<~0.8$.
Here, $\kappa_2(\pr - \apr)/\langle \pr + \apr \rangle$ shows a suppression with respect to the Skellam baseline by roughly 9\% for all multiplicity classes, which is well described by a canonical ensemble as implemented in Thermal-FIST \cite{vovchenko_thermal-fist_2019}.
The optimal correlation volume is found to be $V_\mathrm{c}~=~2.78~\mathrm{d}V/\mathrm{d}y$.
The PYTHIA event generator \cite{bierlich_comprehensive_2022} fails to describe the data since the Lund string fragmentation that it implements happens on length scales of one unit of rapidity.
From the fourth order, volume fluctuations have to be taken into account to describe the data \cite{braun-munzinger_bridging_2017}.
To this end, cumulants of the volume fluctuation contributions \cite{skokov_volume_2013} are estimated using single-particle cumulants of the proton multiplicity from both the data and the model as $\tilde{v}_2 = \frac{1}{\tilde{\kappa}_1} \left[ \left( \frac{\kappa_2}{\kappa_1} \right)_\mathrm{data} - \frac{\tilde{\kappa}_2}{\tilde{\kappa}_1} \right]$ and $\tilde{v}_3 = \frac{1}{\tilde{\kappa}_1^2} \left[ \left( \frac{\kappa_3}{\kappa_1} \right)_\mathrm{data} - \frac{\tilde{\kappa}_3}{\tilde{\kappa}_1} - 3 \tilde{\kappa_2} \tilde{v}_2 \right]$.
Here, $\tilde{\kappa}_n$ signify the model prediction without volume fluctuations included.
They are then added to the model predictions for the cumulant ratios with the volume fluctuation contribution estimation included using

\smallskip
\noindent\begin{minipage}{.35\linewidth}
\begin{equation}
  \frac{\kappa_4}{\kappa_2} = \frac{\tilde{\kappa}_4}{\tilde{\kappa}_2} + 3 {\tilde{\kappa}_2} \tilde{v}_2
\end{equation}
\end{minipage}\hfill%
and%
\hfill
\begin{minipage}{.45\linewidth}
\begin{equation}
  \frac{\kappa_6}{\kappa_2} = \frac{\tilde{\kappa}_6}{\tilde{\kappa}_2} + 15 \tilde{\kappa}_4\tilde{v}_2 + 3 \tilde{\kappa}_4^2\,\tilde{v}_3.
\end{equation}
\end{minipage}
\smallskip

The canonical ensemble with the correlation volume obtained from $\kappa_2(\pr - \apr)/\langle \pr + \apr \rangle$ and with volume fluctuation contributions then describes both fourth- and sixth-order cumulant ratios over the whole multiplicity range.
Therefore, no sign of criticality is observed in pp collisions at $\sqrts = \SI{13}{\TeV}$ up to $\langle \mathrm{d}N_\mathrm{ch}/\mathrm{d}\eta \rangle \approx 31$.

\section{Net-proton fluctuations in Pb--Pb collisions}
\label{pbpbcollisions}

\begin{figure}[b]
    \centering
    \begin{subfigure}{0.333\textwidth}
        \centering
        \includegraphics[width=\textwidth]{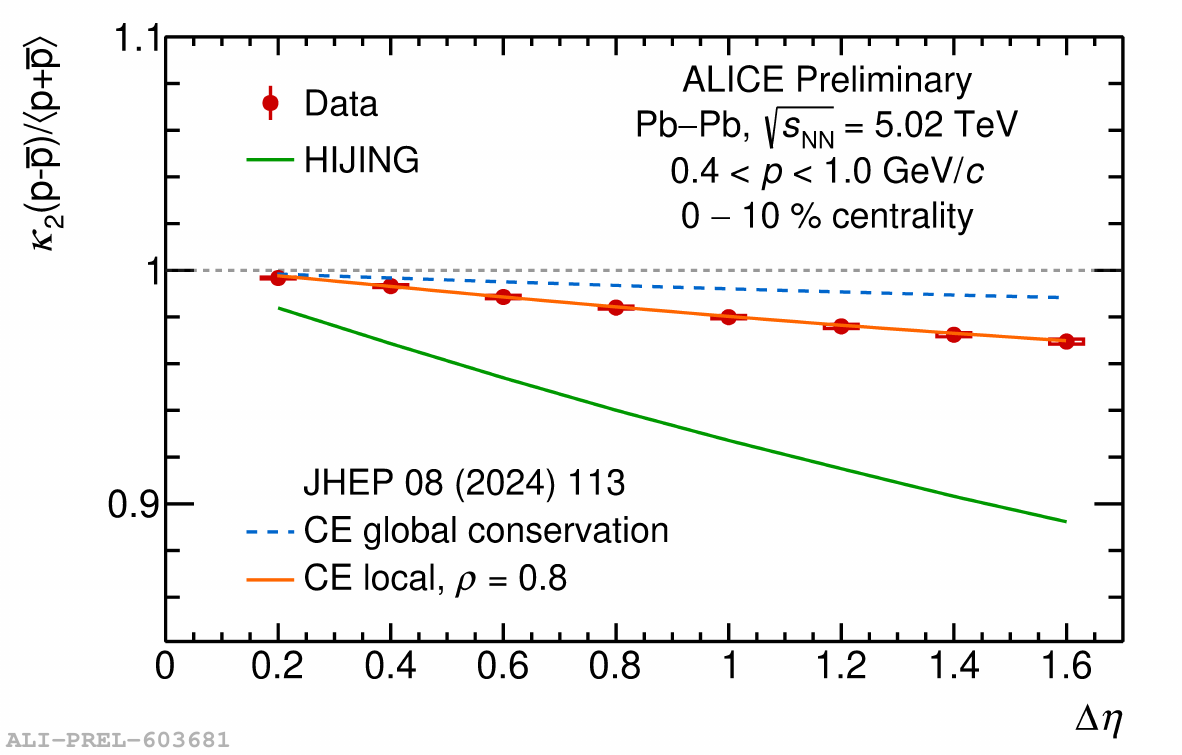}
        \label{fig:pp_k2s_eta}
    \end{subfigure}%
    \begin{subfigure}{0.333\textwidth}
        \centering
        \includegraphics[width=\textwidth]{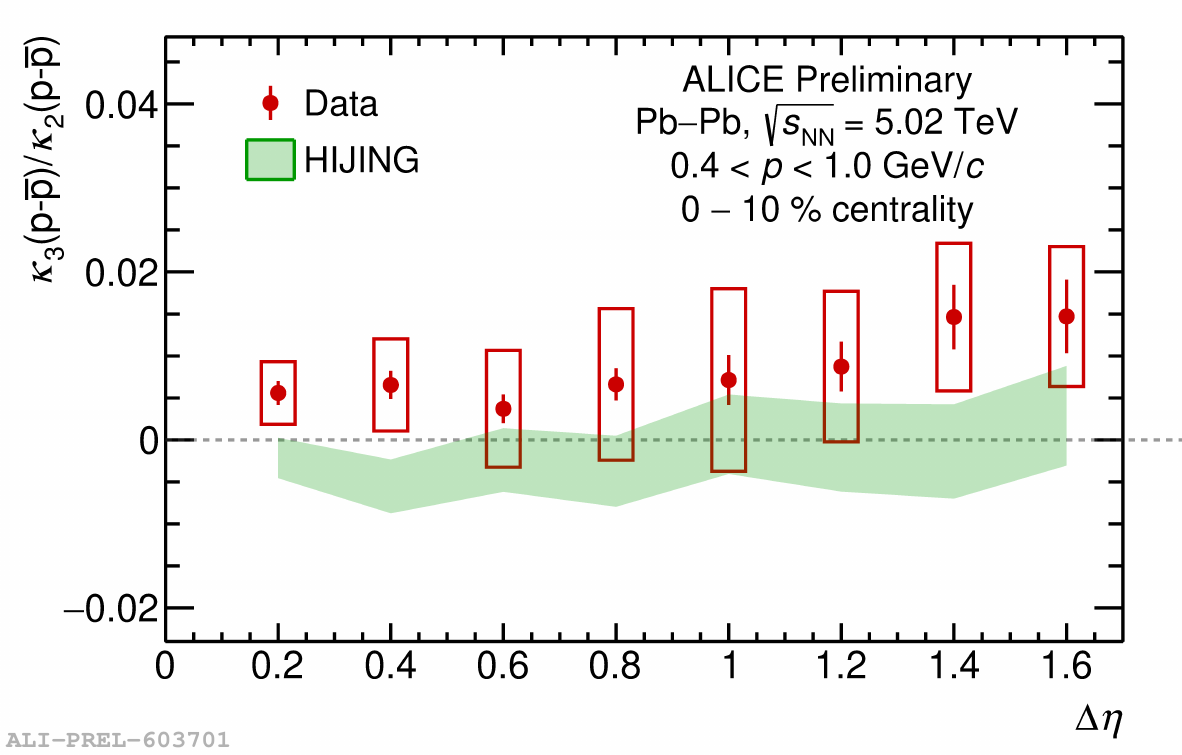}
        \label{fig:pp_k3k2_eta}
    \end{subfigure}%
    \begin{subfigure}{0.333\textwidth}
        \centering
        \includegraphics[width=\textwidth]{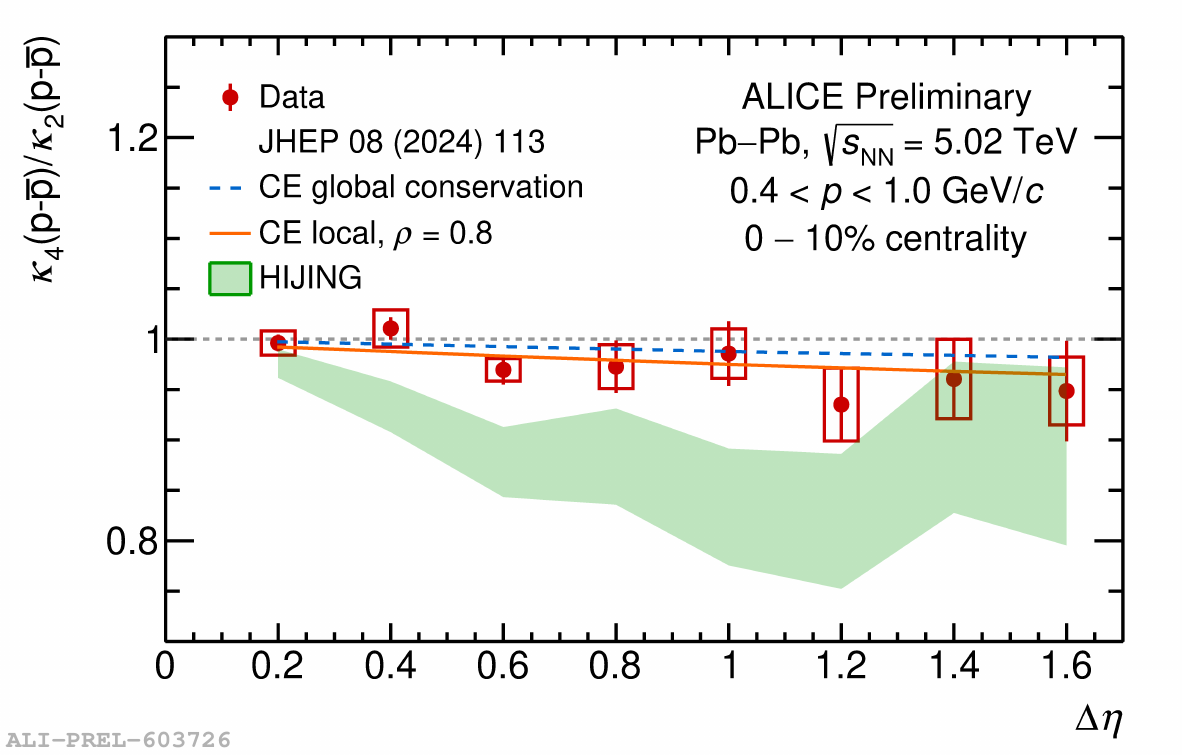}
        \label{fig:pbpb_k4k2_eta}
    \end{subfigure}
    \vspace{-0.5cm}
    \begin{subfigure}{0.333\textwidth}
        \centering
        \includegraphics[width=\textwidth]{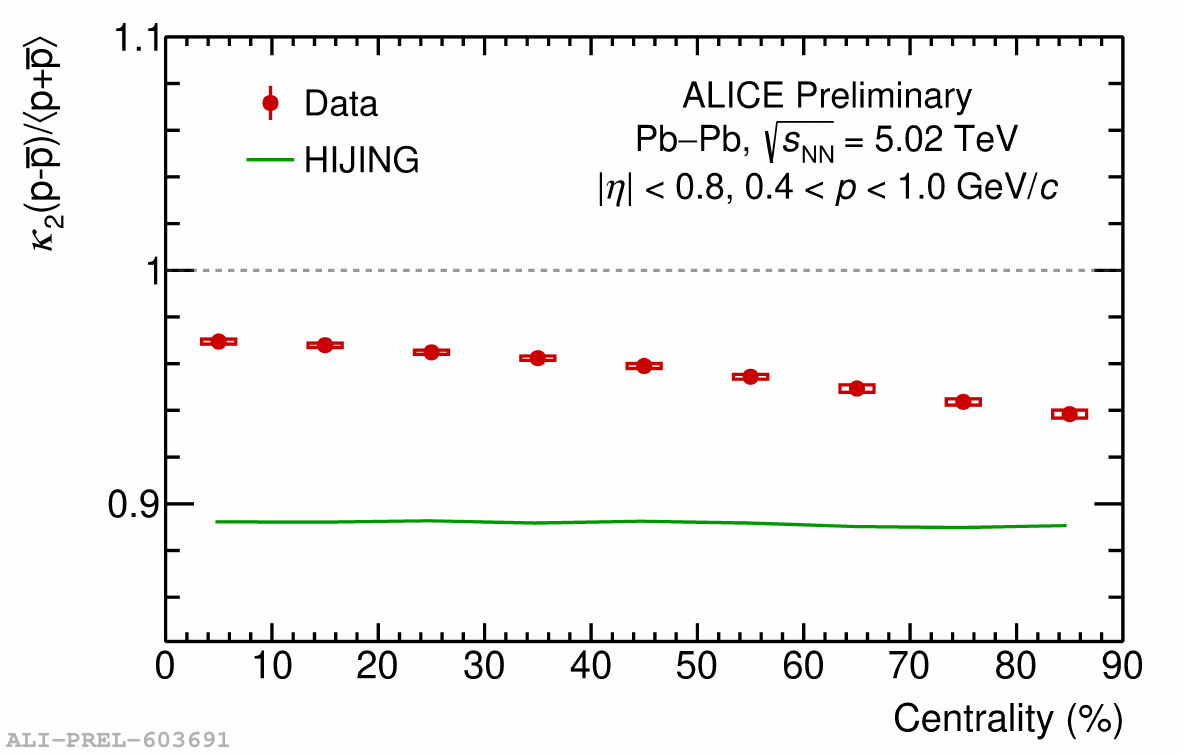}
        \label{fig:pbpb_k2s_cent}
    \end{subfigure}%
    \begin{subfigure}{0.333\textwidth}
        \centering
        \includegraphics[width=\textwidth]{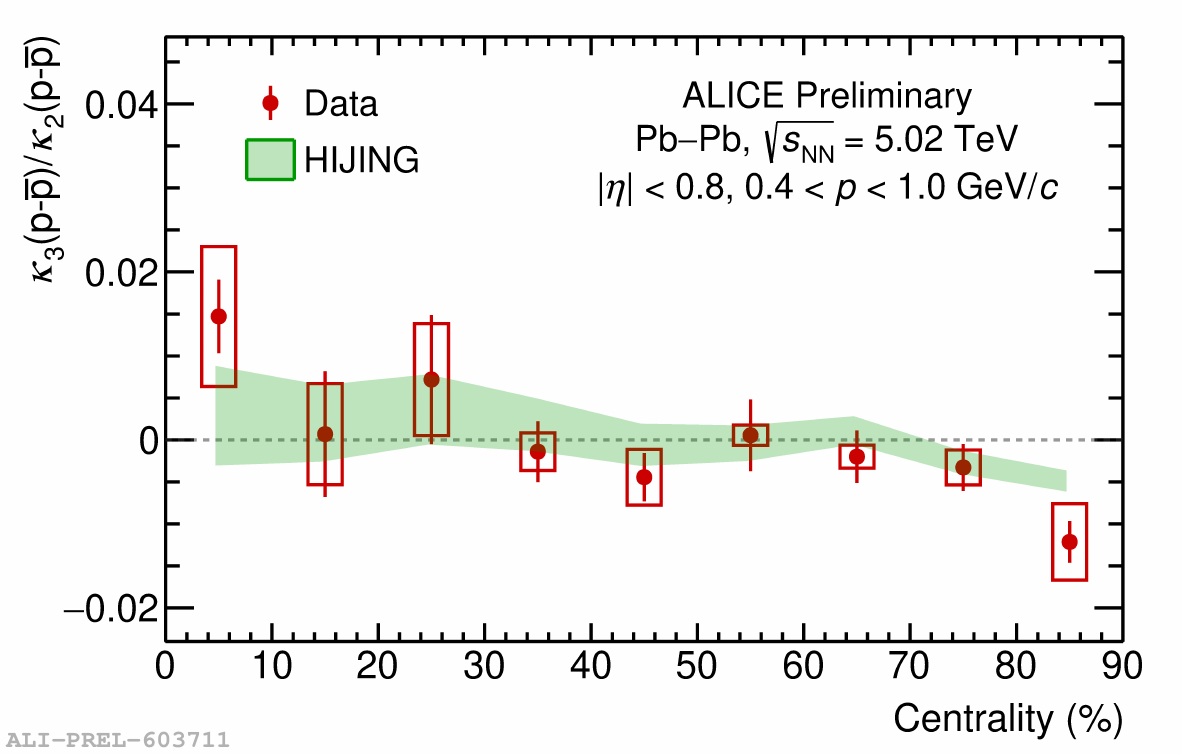}
        \label{fig:pbpb_k3k2_cent}
    \end{subfigure}%
    \begin{subfigure}{0.333\textwidth}
        \centering
        \includegraphics[width=\textwidth]{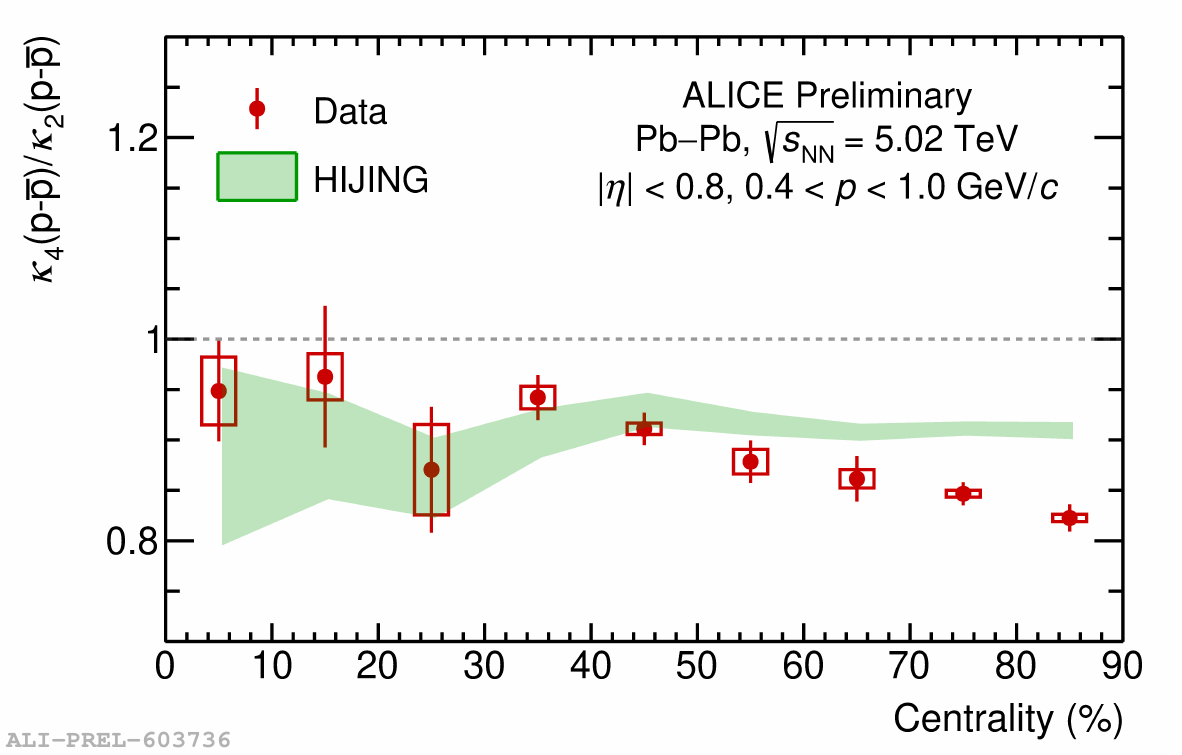}
        \label{fig:pbpb_k4k2_cent}
    \end{subfigure}

    \caption{Net-proton number cumulant ratios $\mathrm{\kappa}_2/\langle~\mathrm{p}~+~\overline{\mathrm{p}}~\rangle$ (left), \kthreektwo{} (center) and \kfourktwo{} (right) in Pb--Pb collisions at $\SI{5.02}{\TeV}$ as a function of the pseudorapidity acceptance $\Delta \eta$ (top) and centrality (bottom).}
    \label{fig:pbpb_cumulants}
\end{figure}

The results from Pb--Pb collisions are shown in \cref{fig:pbpb_cumulants}.
Protons are measured in the smaller momentum range $\SI{0.4}{\GeVc} < p < \SI{1.0}{\GeVc}$ to avoid using the TOF to keep the average efficiency high and the statistical uncertainties low.
The second-order cumulant ratio \ktwos{} shows a suppression with respect to the Skellam baseline between $6\%$ and $1.5\%$ depending on the centrality, which can be explained by a combination of the effect of baryon number conservation and the change in the spectral shape from central to peripheral collisions.
The dependence on the pseudorapidity acceptance can be explained by long range baryon conservation using a canonical ensemble with a correlation coefficient in rapidity $\rho = 0.8$, which corresponds to a correlation length $\Delta y = 5.6$.
The \kthreektwo{} is consistent with zero due to the vanishing $\mu_B$ at LHC energies \cite{alice_collaboration_measurements_2024}.

For \kfourktwo{}, a volume fluctuation correction is applied based on the event mixing technique \cite{rustamov_model-free_2023}.
Collisions in each centrality class are categorized based on the charged particle multiplicity, the event vertex $z$ position and the event plane angle $\Psi_2$.
For each selected collision, a new one is constructed using tracks from other events in the same category, eliminating correlations between particles but keeping the contribution from volume fluctuations.
Cumulants from these are used to calculate the corrected cumulant ratios are shown in \cref{fig:pbpb_cumulants}.
The $\Delta \eta$ dependence in the most central collisions can be also described using $\Delta y = 5.6$, showing that long range baryon number conservation is sufficient to describe the data.
Due to the smaller momentum acceptance and thus reduced proton multiplicity, the deviation from the Skellam baseline is only up 5\% with similarly sized uncertainties.
To better test the LQCD prediction, the acceptance must be enlarged while keeping the efficiency high \cite{braun-munzinger_imprint_2024}.
To this end, a probabilistic method of measuring the moments of multiplicity distributions \cite{rustamov_fuzzy_2024} can be used in the future.

Similarly to PYTHIA, the HIJING event generator \cite{gyulassy_hijing_1994} gives smaller values of $\mathrm{\kappa}_2/\langle~\mathrm{p}~+~\overline{\mathrm{p}}~\rangle$ than the data, which can be explained by the Lund string fragmentation, which results in shorter range correlations.
The data from HIJING simulations are corrected using mixed events as in data for the comparison of fourth-order cumulants.
The apparent agreement in the ratio should not be taken seriously because $\kappa_2(\pr - \apr)$ in the denominator is badly described.

\section{Conclusions}
\label{conclusions}

Ratios of net-proton cumulants in pp and Pb--Pb collisions at the LHC have been presented.
They can be described using correlations stemming from local baryon number conservation without any critical behaviour.
Probabilistic analysis techniques can be used to enlarge the momentum acceptance to test LQCD predictions.
With the particle identification capabilities in large acceptance of the next generation detector ALICE 3, it will be possible to extend the measurement in heavy-ion collisions to sixth and higher orders.


\bibliography{ebye2.bib}


\end{document}